\documentclass[10pt,twocolumn]{article}
\usepackage{amsfonts}
\usepackage{amsmath,graphicx,fancyhdr}
\usepackage{mathrsfs}
\usepackage{indentfirst}
\usepackage{geometry}
\usepackage{times}
\usepackage{simpleConference}
\usepackage{cite}
\usepackage{setspace}
\usepackage{caption}

\makeatletter
\newdimen\bibspace
\renewenvironment{thebibliography}[1]
{\section*{\refname}%
\@mkboth{\MakeUppercase\refname}{\MakeUppercase\refname}%
\list{\@biblabel{\@arabic\c@enumiv}}%
{\settowidth\labelwidth{\@biblabel{#1}}%
\leftmargin\labelwidth
\advance\leftmargin\labelsep
\itemsep\bibspace
\parsep\z@skip
\@openbib@code
\usecounter{enumiv}%
\let\p@enumiv\@empty
\renewcommand\theenumiv{\@arabic\c@enumiv}}%
\sloppy
\clubpenalty4000
\@clubpenalty \clubpenalty
\widowpenalty4000%
\sfcode`\.\@m}
{\def\@noitemerr
{\@latex@warning{Empty `the bibliography' environment}}%
\endlist}
\renewcommand\@biblabel[1]{{[#1]\hfill}}
\makeatother

\renewcommand\thesection{\arabic{section}.\kern -.2em}
\newcommand{\xiaoer}{\fontsize{18pt}{18pt}\selectfont}

\newcommand{\xiaosi}{\fontsize{12pt}{12pt}\selectfont}

\newcommand{\wuhao}{\fontsize{10pt}{10pt}\selectfont}

\def\abstract
   {%
   \leftline{\xiaosi\bf Abstract}%
   \vspace*{8pt}%
   \upshape%
   }

\def\vm#1{\mathrm{\bf #1}}

\begin{document}

\title{\xiaoer\ PRACTICAL DESIGN OF MULTI-CHANNEL OVERSAMPLED 
WARPED COSINE-MODULATED FILTER BANKS }
\author{\xiaosi\ \textbf{M.I. Vashkevich*, W. Wan$^\dag$, A.A. Petrovsky*} 
\\ \wuhao\ *Belarusian State University of Informatics and Radioelectronics,  Computer Egeneering Department, 
\\ \wuhao\ st. P.Brovky, 6, 220013, Minsk, Belarus, email: \{vashkevich, palex\}@.bsuir.by 
\\ \wuhao\ $^\dag$Shanghai University,  School of Communication and Information Engineering,
\\ \wuhao\ Shanghai 200072, P.R. China, email: wanwg@staff.shu.edu.cn}
\date{}
\maketitle
\pagestyle{plain}
\setcounter{page}{1}
\cfoot{\thepage\ }
\rfoot{CCWMC2011}
\renewcommand{\headrulewidth}{0pt}
\renewcommand{\footrulewidth}{0pt}

\thispagestyle{fancy}
\begin{Keywords} 
	Nonuniform filter bank, optimization.
\end{Keywords}

\begin{abstract}
A practical approach to optimal design of multichannel oversampled warped cosine-modulated filter banks (CMFB) is proposed. Warped CMFB is obtained by allpass transformation of uniform CMFB. The paper addresses the problems of minimization amplitude distortion and suppression of aliasing components emerged due to oversampling of filter bank channel signals. Proposed optimization-based design considerably reduces distortions of overall filter bank transfer function taking into account channel subsampling ratios. Matlab-implementation of proposed warped CMFB design method is available in public GitHub repository.
\end{abstract}
\section{\xiaosi\ \textbf{Introduction}}
\label{sec:intro}

Multirate filter banks are widely used as computationally efficient and flexible building blocks for subband signal processing~\cite{01}. These filter banks decompose an input signal into its subband components and reconstruct the signal from the downsampled version of these components with little or no distortion.

In various applications such as noise reduction, speech enhancement and audio coding nonuniform time-fre-quency representation is highly desired. A well-known example is approximation of critical bands of human auditory system. By using nonuniform filter banks this problem can be effectively solved. Another problem that is solved by means of nonuniform filter banks is an estimation of the frequency dependent reverberation time~\cite{10}.

One simple way of obtaining nonuniform filter bank is employing allpass transform to uniform filter bank~\cite{02}. Efficient nonuniform DFT polyphase filter banks were proposed in~\cite{03,04}. However, DFT-based filter banks produce complex-value channel signals even for real-value input. Therefore the subsequent subband processing becomes more sophisticated. In contrast to DFT filter banks allpass transformed (or warped) cosine-modulated filter banks (CMFB) developed in~\cite{05} are allow to avoid having complex subband signals for a real-value input. It is worth to mention that aliasing cancellation conditions~\cite{06} do not hold for warped CMFB thus all aliasing components should be suppressed by synthesis filter bank.

The paper presents a practical approach to optimal design of multichannel oversampled warped CMFB with low aliasing and amplitude distortions. Just like uniform CMFB, analysis and synthesis filters of warped CMFB are obtained from one prototype filter, that results in high design efficiency. A distinguishing characteristic of warped CMFB is that for each channel subsampling factors should be determined separately~\cite{07}. The practical rule of selection subsampling factors is also derived.

\section{\xiaosi\ \textbf{Warped cosine-modulated filter bank}}
\subsection{\wuhao\ \textbf{Uniform CMFB: brief review}}
The $M$-channel warped CMFB proposed in~\cite{05} based on uniform CMFB developed in~\cite{06}. The impulse responses of the analysis ($h_k[n]$) and synthesis ($f_k[n]$) filters are cosine-modulated versions of the prototype filter $h[n]$:
\begin{subequations}
\begin{equation*}
\textstyle
h_k[n]=2h[n]\cos\left[ \frac{(2k+1)\pi}{2M}(n-\frac{N-1}{2})+\theta_k)\right], 
\end{equation*}
\begin{equation*}
\textstyle
f_k[n]=2h[n]\cos\left[ \frac{(2k+1)\pi}{2M}(n-\frac{N-1}{2})+\phi_k)\right], 
\end{equation*}
\end{subequations}
where $0\leq n < N$,  $0\leq k < M$ and
$ \theta_k = -\phi_k = (-1)^k \frac{\pi}{4}$.

The length of $h[n]$ is assumed to be multiple of $2M$, i.e. $N=2mM$. The transfer functions of analysis and synthesis filters can be expressed as follows:
\begin{subequations}
\label{fb_z_trans}
\begin{equation}
\textstyle
H_k(z)=a_k b_k H(z W_{2M}^{k+0.5}) + a_k^* b_k^* H(z W_{2M}^{-(k+0.5)}),
\end{equation}
\begin{equation}
\textstyle
F_k(z)=a_k^* b_k F(z W_{2M}^{k+0.5}) + a_k b_k^* F(z W_{2M}^{-(k+0.5)}), 
\end{equation}
\end{subequations}
where $a_k = e^{j(-1)^k\frac{\pi}{4}}$, $b_k = W_{2M}^{(k+0.5)\frac{N-1}{2}}$ and $W_{2M}=e^{-j\pi/M}$. The superscript $*$ denotes the complex conjugation. In (\ref{fb_z_trans}) $H(z)$ is linear phase lowpass FIR filter prototype with cutoff frequency $\omega_c\approx \pi/M$.

\subsection{\wuhao\ \textbf{Allpass transformed CMFB}}
\begin{figure}[htb]
\centering
\includegraphics[width=80mm]{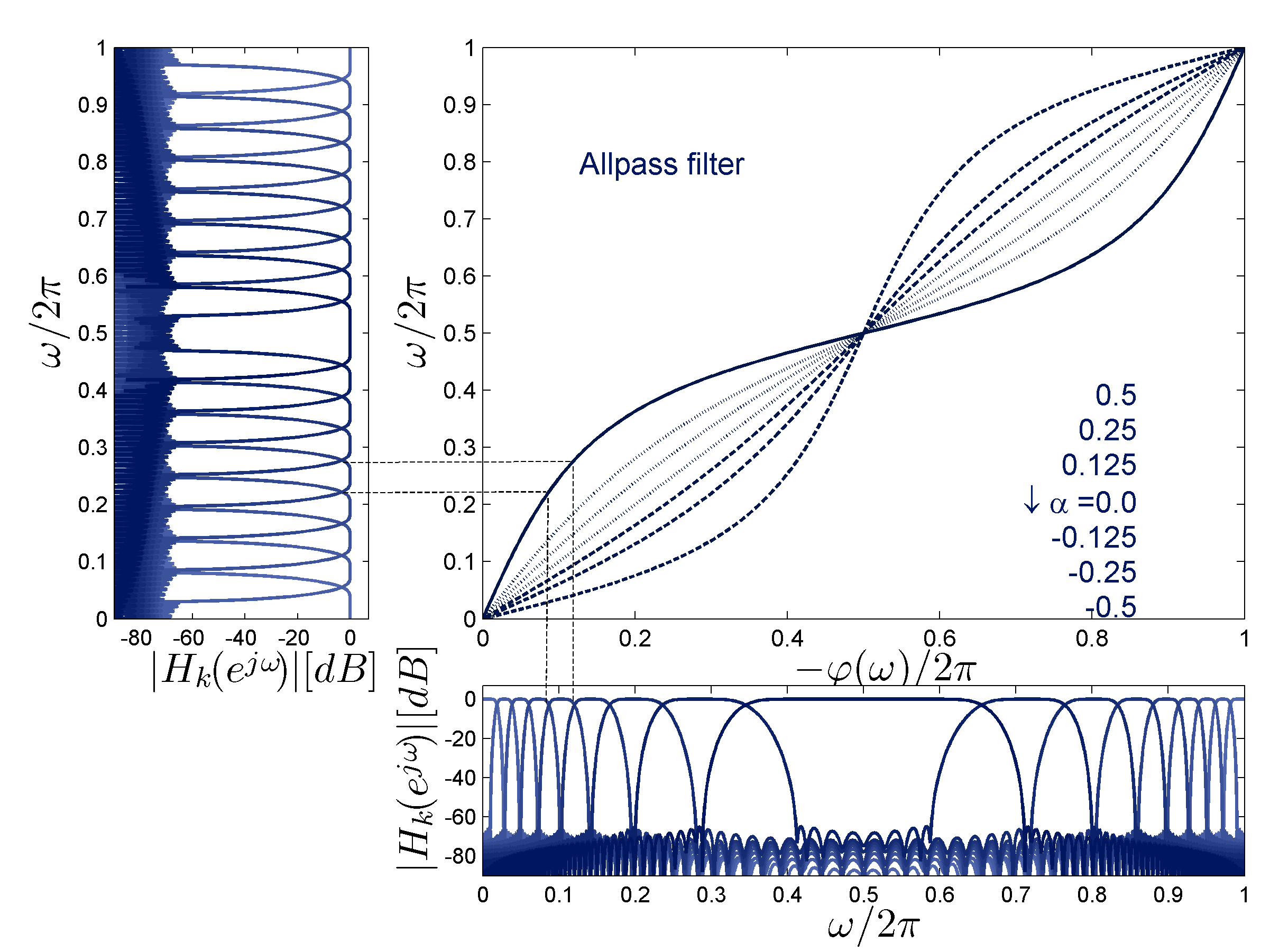}
\caption{Allpass transformation of CMFB}
\label{fig1}
\end{figure}

Allpass transformation of CMFB consists in replacing all the delays in uniform CMFB with causal and stable allpass filters:
\begin{equation}
z^{-1}\rightarrow A(z).\label{allpass_tr}
\end{equation}
In this paper we consider first-order allpass filter
\begin{equation*}
A(z)=\frac{z^{-1}+\alpha}{1+\alpha z^{-1}},\quad \alpha\in \mathbb{R}, \quad |\alpha|<1,
\end{equation*}
with frequency response $A(e^{j\omega})=e^{j\varphi(\omega)}$. The phase response is written as
\begin{equation}
\varphi(\omega)=-\omega+2\arctan\left(\frac{\alpha\sin\omega}{\alpha\cos\omega - 1} \right). \label{ap_transf}
\end{equation}
Thus, replacing all terms  $z^{-1}$ by by first order allpass filters leads to a transformation $\omega\rightarrow\varphi(\omega)$ of the frequency scale as shown in figure~\ref{fig1}.

\begin{figure}[htb]
\centering
\includegraphics[width=80mm]{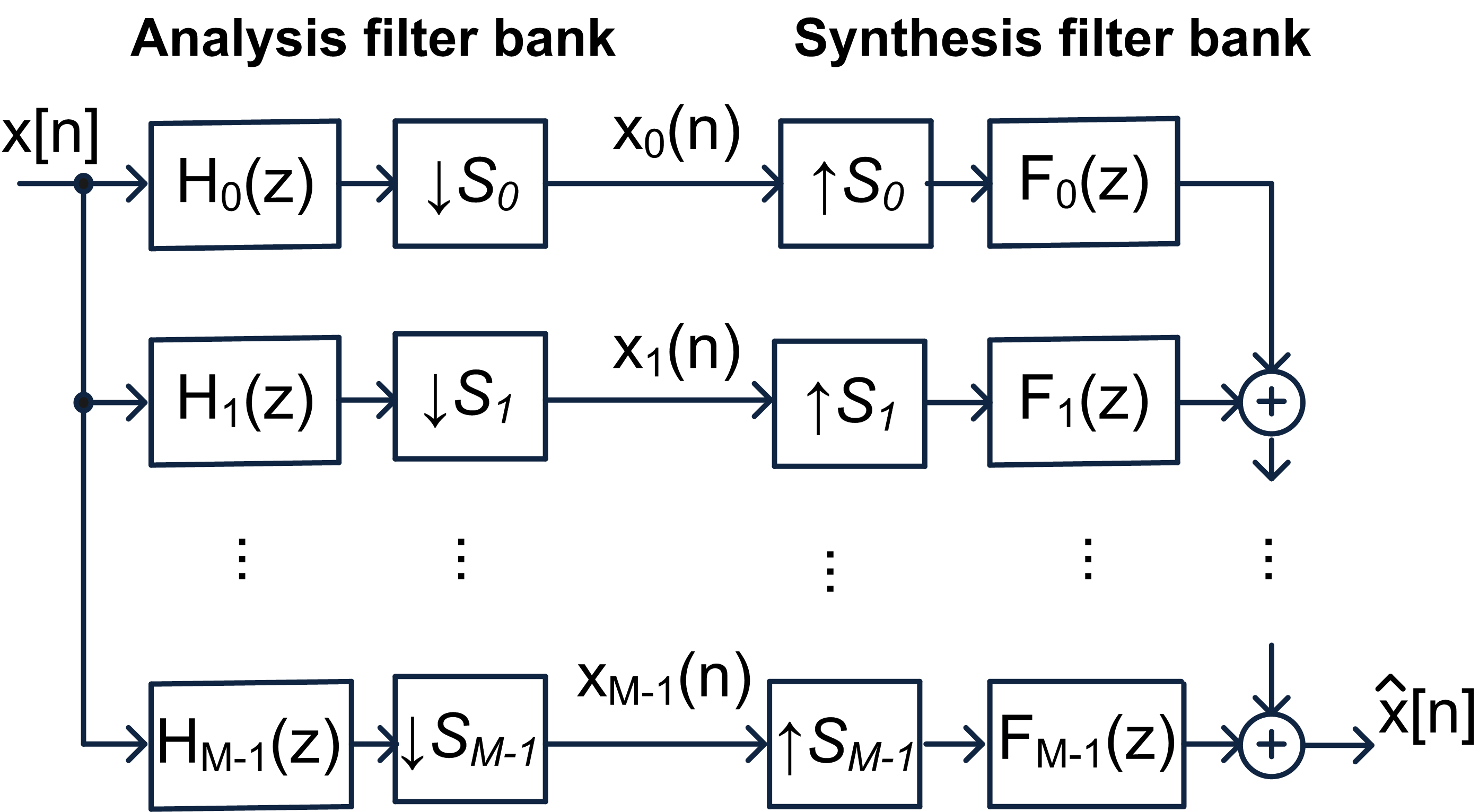}
\caption{Allpass transformation of CMFB}
\label{fig2}
\end{figure}

\section{\xiaosi\ \textbf{Problem statement}}
Figure~\ref{fig2} shows the block diagram of the nonuniform oversampled filter with overall transfer function 
\begin{equation}\textstyle
T_{all}(z) = T_{dist}(z)+T_{alias}(z), \label{T_all}
\end{equation}
where
\[  T_{dist}(z) = \sum_{k=0}^{M-1}H_k(z)F_k(z), \]
\[  T_{alias}(z)= \sum_{k=0}^{M-1}\sum_{l=0}^{S_k-1} H_k(zW_{S_k}^l)F_k(z).\]
Distortion transfer function $T_{dist}(z)$ shows amplitude distortion introduced by analysis and synthesis filters, while aliasing distortion is described by an aliasing transfer function $T_{alias}(z)$. In perfect reconstruction case $T_{alias}(e^{j\omega})=0$ and $|T_{dist}(e^{j\omega})|=1, \forall \omega \in [0,\,\pi]$. Thus optimization problem for the design of the oversampled warped CMFB is formulated as follows 
\begin{equation}
\label{err_fun}\textstyle
\min |E(\omega)|, \quad \forall \omega \in [0,\,\pi],
\end{equation}
where the error function $E(\omega)$ is defined as
\begin{equation}\textstyle
E(\omega) = |T_{all}(e^{j\omega})|^2 - 1, \quad \omega\in [0,\,\pi].
\label{eq:expr_error_fun}
\end{equation}
A weighted least square objective function can be formulated to solve optimization problem~(\ref{err_fun}),
\begin{equation}
g=\sum_{\omega \in [-0,\;\pi]} B(\omega)|E(\omega)|^2,
\label{obj_fun}
\end{equation}
where $B(\omega)$ is nonnegative weighting function.

It is difficult to obtain the optimal solution that minimizes~(\ref{obj_fun}). In~\cite{08} it has been shown that an iterative algorithm can be derived for the design of CMFB that jointly optimizes the filter prototype coefficients and weighted function. The design algorithm given in this paper adopts this approach for the minimization of~(\ref{obj_fun}) to archive solution that is near optimal in minimax sense.

\section{\xiaosi\ \textbf{The design algorithm}}
\subsection{\wuhao\ \textbf{Preliminaries}}
It is well known that frequency response of linear phase FIR filter can be written as
\begin{equation*}\textstyle
H(\omega)=e^{-j[(N-1)\omega/2]} \vm{C}^T(\omega)\vm{h},
\end{equation*}
where 
\begin{equation} \notag
\textstyle \vm{C}(\omega)=
\begin{cases}
\left[2\cos\left(\frac{\omega}{2}\right)\ 2\cos\left(\frac{3\omega}{2}\right) \ldots 2\cos\left(\frac{[N-1]\omega}{2}\right)\right]^T\\ \text{$N$ even}\\
\left[1\ 2\cos\left(\omega \right)\ldots 2\cos\left(\frac{[N-1]\omega}{2}\right) \right]^T \\ \text{$N$ odd}
\end{cases}
\end{equation}
\begin{equation} \textstyle\notag
\vm{h}=\left[h\left( \bigl\lfloor \frac{N}2 \bigr\rfloor\right) \> h\left( \bigl\lfloor \frac{N}2 \bigr\rfloor+1\right) \ldots h\left(N-1 \right) \right].
\end{equation}
The operator $\lfloor a \rfloor$ denotes the largest integer smaller than $a$.
The superscript $^T$ indicate matrix transpose. Thus the overall transfer function (\ref{T_all}) can be expressed as follows
\begin{equation}\textstyle
T_{all}(e^{j\omega})=\vm{h}^T \vm{U}(\omega) \vm{h}, \label{T_all2}
\end{equation}
where
\begin{equation*}
\begin{split} \textstyle
\vm{U}(\omega) &= \\
\sum_{k=0}^{M-1}& \sum_{l=0}^{S_k-1} \bigl[  a_k b_k e^{-j[(N-1)\gamma_1(\omega,l,k)/2])} \vm{C}^T(\gamma_1(\omega,l,k))\\ 
+& a_k^* b_k^* e^{-j[(N-1)\gamma_2(\omega,l,k)/2])} \vm{C}^T(\gamma_2(\omega,l,k)) \bigr] \times \\
\bigl[ & a_k^* b_k e^{-j[(N-1)\gamma_1(\omega,0,k)/2])} \vm{C}^T(\gamma_1(\omega,0,k))\\ 
+& a_k b_k^* e^{-j[(N-1)\gamma_2(\omega,0,k)/2])} \vm{C}^T(\gamma_2(\omega,l,k)) \bigr]^T
\end{split}
\end{equation*}
and
\[ \textstyle \gamma_1(\omega,l,k)=-\phi(\omega+2\pi l/S_k)-2\pi(k+0.5)/2M, \]
\[ \textstyle \gamma_2(\omega,l,k)=-\phi(\omega+2\pi l/S_k)+2\pi(k+0.5)/2M. \]

\subsection{\wuhao\ \textbf{Optimization procedure}}
The design algorithm consists of two iterative procedures: one nested within the other. The vector of coefficients $\vm{h}_{\ell}$ are iteratively optimized in inner loop procedure. In the outer loop procedure the weighted function $B_\mu (\omega)$ is updated. During inner loop procedure $B_\mu (\omega)$ is kept constant. Subscripts $\ell$ and $\mu$ are used to indicate values at the $\ell$-th iteration of inner loop procedure and $\mu$-th iteration of outer procedure, respectively.

Using (\ref{T_all2}) the objective function (\ref{obj_fun}) at the $\ell$-th iteration can be expressed as
\begin{equation}
g(\vm{h}_{\ell})=\sum_{\omega \in [-0,\;\pi]} B_{\mu}(\omega)|E_\ell(\omega)|^2,
\label{obj_fun_2}
\end{equation}
where
\[ \textstyle E_\ell(\omega) = \bigl[ \vm{h}_\ell^T\vm{U}_r(\omega)\vm{h}_\ell \bigr]^2 + \bigl[ \vm{h}_\ell^T\vm{U}_i(\omega)\vm{h}_\ell \bigr]^2 - 1,
\]
\[ \textstyle
\vm{U}_r(\omega) = \mathop{\text{Re}} (\vm{U}(\omega)), \quad \vm{U}_i(\omega) = \mathop{\text{Im}} (\vm{U}(\omega)).
\]

The necessary condition for minimization of (\ref{obj_fun}) is equality of gradient $\bigtriangledown g$ to zero. Differentiating (\ref{obj_fun_2}) with respect to $\vm{h}$ 
\begin{equation} \label{grd}
\bigtriangledown g(\vm{h}_{\ell})=2\sum_{\omega\in[0,\,\pi]} B_\mu(\omega)E_\ell(\omega) \cdot \mathrm{grad}(E_\ell(\omega)),
\end{equation}
where
\begin{equation} \notag
\begin{split} \textstyle
\mathrm{grad}&(E_\ell(\omega)) = \\
 2\Bigl(&\vm{h}_\ell^T \vm{U}_r(\omega) \vm{h}_\ell (\vm{U}_r(\omega) + \vm{U}^T_r(\omega)) \vm{h}_\ell + \\
& \vm{h}_\ell^T \vm{U}_i(\omega) \vm{h}_\ell (\vm{U}_i(\omega) + \vm{U}_i^T(\omega)) \vm{h}_\ell\Bigr).
\end{split}
\end{equation}

Let $\vm{h}_{\ell,opt}=\vm{h}_\ell + \vm{e}_\ell$ denotes the optimum solution that sets (\ref{grd}) to zero. Expanding $\bigtriangledown g$ in Taylor's series
\begin{equation} \textstyle
\bigtriangledown g (\vm{h}_\ell + \vm{e}_\ell) = \bigtriangledown g(\vm{h}_\ell) + \bigtriangledown^2 g(\vm{h}_\ell)\cdot \vm{e_\ell}+\ldots = 0. \label{tayl}
\end{equation}
Without taking into consideration higher order term in~(\ref{tayl}) the correction vector $\vm{e}_\ell$ can be obtained by solving
\begin{equation} \textstyle
\vm{e}_l=-[\bigtriangledown^2 g(\vm{h}_\ell)]^{-1} \cdot
\bigtriangledown g(\vm{h}_\ell).
\label{cor_vect_eq}
\end{equation}
$\bigtriangledown^2 g$ is hessian of objective function that is given by
\begin{equation}
\begin{split}
\bigtriangledown^2 g(\vm{h}_\ell) = &2 \sum_{\omega\in[0,\;\pi]}   B_\mu(\omega) \bigl[ \mathrm{grad}(E_\ell(\omega))\times \\ &\mathrm{grad}(E_\ell(\omega))^T + E_\ell(\omega)\times \vm{G}(\omega) \bigr], 
\end{split}
\end{equation}
where
\begin{equation}\notag
\begin{split}
\vm{G}(\omega) = & 2 \Bigl( \vm{h}_\ell(\vm{U}_r(\omega)+\vm{U}_r^T(\omega))^T(\vm{U}_r(\omega)+\vm{U}_r^T(\omega))\vm{h}_\ell\\ &+\vm{h}_\ell^T \vm{U}_r(\omega) \vm{h}_\ell (\vm{U}_r(\omega) + \vm{U}_r^T(\omega)) \\
&+\vm{h}_\ell(\vm{U}_i(\omega)+\vm{U}_i^T(\omega))^T(\vm{U}_i(\omega)+\vm{U}_i^T(\omega))\vm{h}_\ell\\
&+\vm{h}_\ell^T \vm{U}_i(\omega) \vm{h}_\ell (\vm{U}_i(\omega) + \vm{U}_i^T(\omega))\Bigr)
\end{split}
\end{equation}
Iterative procedure starts with $\vm{h}_0$, which is a rough estimation of optimum, that can be obtain using available FIR filter design method (windowing, etc.). At the $\ell$-th iteration, the correction vector $\vm{e}_\ell$ obtained by solving~(\ref{cor_vect_eq}). Then coefficient vector is updated using
\[ \textstyle \vm{h}_{\ell+1}=\vm{h}_\ell+\vm{e}_\ell,  \]
inner loop procedure terminates when $||\vm{e}_l||^2\le 10^{-10}.$

The outer loop procedure updates $B_{\mu}(\omega)$ as described below. Let $\vm{h}_{\mu, opt}$ denote the optimum solution obtained by inner loop procedure for weighting function $B_{\mu}(\omega)$. The error function $E_{\mu}(\omega)$ is obtained using (\ref{eq:expr_error_fun}) and $\vm{h}_{\mu, opt}$. Let $V_\mu (\omega_l)$ be the $l$-th extremum of $|E_{\mu}(\omega)|$. If $V_{\mu}(\omega_l)\leq \min [V_{\mu}(\omega_{l-1}),\, V_{\mu}(\omega_{l+1})]$, then let $V_{\mu}(\omega_l)$ equal this threshold. Define the envelope function $\beta_\mu(\omega),\,\omega\in [0,\,\pi]$ as
\begin{equation*} \textstyle
\beta_\mu(\omega) = \frac{\omega - \omega_l}{\omega_{l+1}-\omega_l}V_\mu(\omega_{l+1}) + \frac{\omega_{l+1} - \omega}{\omega_{l+1}-\omega_l}V_\mu(\omega_l),
\end{equation*}
where $l$ is chosen such that $\omega_{l}\leq\omega\leq\omega_{l+1}$. This envelope function shows that a larger relative weight should be assigned at 
$\omega$ where $|E_{\mu}(\omega)|$ is large.

The new weighting function is obtained as
\begin{equation*} \textstyle
B_{\mu+1} (\omega) = B_{\mu}(\omega)\frac{\beta_\mu(\omega)^\theta}{A_\mu},
\end{equation*}
where the normalization factor $A_\mu$ is defined as
\[
A_\mu=\sqrt{\sum_{\omega\in[0,\,\pi]}\beta_\mu(\omega)^{2\theta}},
\]
and $\theta$ is a positive factor that affects the convergence ($1\leq \theta \leq 1.5$). Iterative algorithm terminates when
\[ \textstyle
\frac{\max\beta_\mu(\omega)-\min\beta_\mu(\omega)}{\max\beta_\mu(\omega)+\min\beta_\mu(\omega)} \leq \psi,
\]
where approximate value of $\psi = 0.5 \ldots 0.7$.

\section{\xiaosi\ \textbf{Subsampling ratio selection}}
In~\cite{07} the following rule is given for selection of subsampling ratios for warped CMFB
\begin{equation} \textstyle
\Big\lfloor \frac{n_k}{2f_{U_k}} \Big\rfloor \geq S_k \geq \Big\lceil \frac{n_k-1}{2f_{L_k}} \Big\rceil, \; 1\leq n_k \leq \Big\lfloor \frac{f_{U_k}}{f_{U_k} - f_{L_k}} \Big\rfloor,
\label{eq_old_rule}
\end{equation}
where $f_{U_k}$ $f_{L_k}$ are normalized lower and upper edge of the $k$-th subband.

In the uniform case $f_{L_k}$ and $f_{U_k}$ are match the passband of $k$-th subfilter, that is possible due to mechanism of aliasing cancellation~\cite{06}. However in the warped CMFB aliasing cancellation conditions do not hold. Therefore, the subsampling factors $S_k$ in the $k$-th subband has to be chosen such that the aliasing components in each subband do not overlap with subband spectrum of input signal. Selection of range $[f_{L_k},\,  f_{U_k}]$ should be based on the same principle as the mechanism of aliasing cancellation \emph{nonadjacent channels do not overlap}. This means that the range $[f_{L_k},\,  f_{U_k}]$ should include passbands of $(k-1)$-th $k$-th and $(k+1)$-th subfilters.

\begin{figure}[htb]
\centering
\includegraphics[width=76mm]{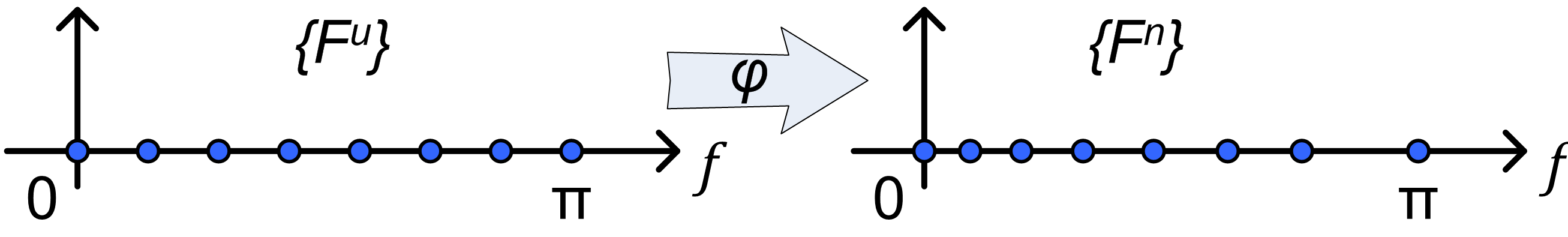}
\caption{Transformation of frequency axis by mapping~(\ref{ap_transf}).}
\label{fig_3}
\end{figure}

The general algorithm for finding the boundaries of the frequency band $[f_{L_k},\,  f_{U_k}]$ defines as follows. The ordered set

\begin{equation}\textstyle
F^u = \left\{ \frac{\pi k}{M} \right\}, \quad k=0,1\ldots M,
\end{equation}
defines passband frequency of subfilters of uniform CMFB. Mapping~(\ref{ap_transf}) allows to obtain corresponding frequency edges of warped CMFB (figure~\ref{fig_3})
\begin{equation*} \textstyle
\varphi\colon F^u \rightarrow F^n.
\end{equation*}

According to proposed rule the frequency band for $k$-th channel of warped CMFB expressed as
\begin{equation} \textstyle
f_{L_k}=\varphi(F_{k-1}^u)/2\pi, \quad f_{U_k} = \varphi(F_{k+2}^u)/2\pi,
\label{eq_newrule1}
\end{equation}
for $k=0$ and $k=M-1$ the following relation holds 
\begin{subequations}
\begin{equation}\textstyle
f_{L_0}=0,\qquad f_{U_0}=\varphi(F^u_{2})/2\pi,
\end{equation}
\begin{equation} \textstyle
f_{L_{M-1}}=\varphi(F^u_{M-2})/2\pi,\; f_{U_{M-1}}=\varphi(F^u_{M})/2\pi,
\end{equation}
\label{eq_new_rule}
\end{subequations}
Joint use of rule (\ref{eq_old_rule})--(\ref{eq_new_rule}) allows determining the subsampling ratios $S_k$ for warped CMFB.

\section{\xiaosi\ \textbf{Design example}}
Let us consider the $22$-channel warped CMFB approximating the psychoacoustic Bark scale for standard sampling frequency $16$ kHz. At first it is necessary to determine warping coefficient $\alpha$. According to~\cite{09} for $f_s=16$ kHz $\alpha=0.5783$. Considering rule described in previous section it is possible to choose the subsampling ratios $S_k$ (figure~\ref{fig2}) such that aliasing does not affect channel signals. For instanse examine the subsampling factor $S_3$. Necessary frequency bounds of uniform CMFB given bellow
\[ \textstyle F_2^u=\frac{2\pi}{22}, \quad F_5^u=\frac{5\pi}{22}.\]

Using (\ref{eq_newrule1}) the corresponding frequency range of warped CMFB can be determined as
\begin{equation} \textstyle
f_{L_3}=\frac{\varphi(F_2^u)}{2\pi}\approx 0.0122, \quad f_{U_3} = \frac{\varphi(F_5^u)}{2\pi}\approx 0.0316,
\end{equation}
The subsampling factor $S_3$ is obtained by applying~(\ref{eq_old_rule})
\[ \textstyle
1\leq n_3 \leq \Big\lfloor \frac{f_{U_3}}{f_{U_3}-f_{L_3}} \Big\rfloor = \lfloor 1.6303\rfloor=1 \Rightarrow n_3=1,
\]  \[ \textstyle 
\Big\lfloor \frac{n_3}{2f_{U_3}} \Big\rfloor \geq S_3 \geq \Big\lceil \frac{n_3-1}{2f_{L_3}} \Big\rceil \Rightarrow \Big\lfloor \frac{1}{0.0632} \Big\rfloor \geq S_3 \geq \Big\lceil \frac{0}{0.0244} \Big\rceil 
\] \[ \textstyle
\Rightarrow \lfloor 15.81 \rfloor \geq S_3 \geq \lceil 0 \rceil \Rightarrow S_3 = 15.
\]
Similarly, the remaining subsampling factors for filter bank under consideration are selected
\begin{multline} \textstyle
S_k = \{ 56,27,20,15,12,21,18,15,\\ 13,12,10,9,7,6,5,5,4,3,1,1,2,3\}.
\end{multline}

\begin{figure}[h]
\centering
\includegraphics[width=80mm]{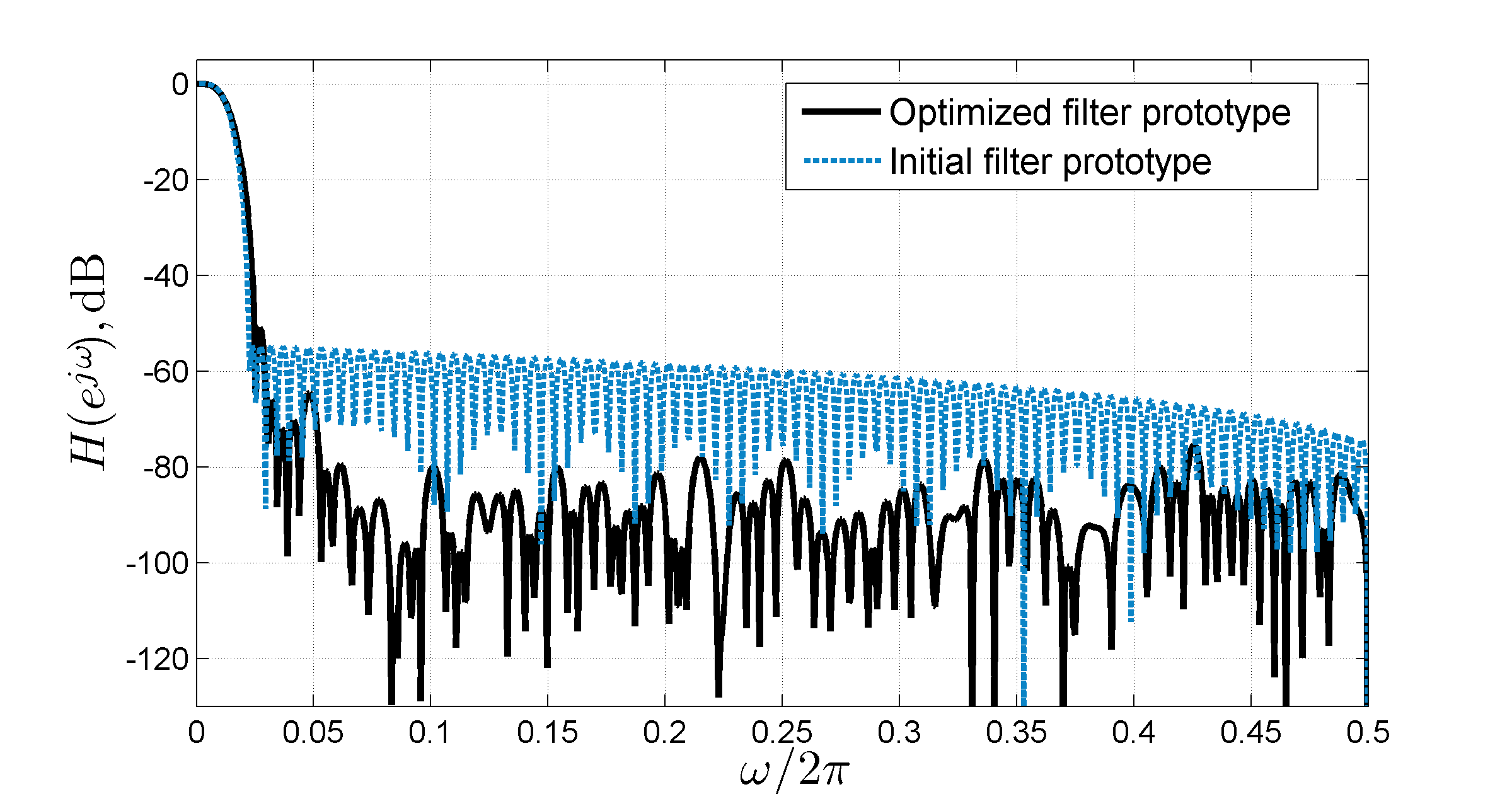}
\caption{Magnitude responses of initial (using~\cite{08}) and optimized filter prototypes of order $N=176$.}
\label{fig_4}
\end{figure}

\begin{figure}[h]
\centering
\includegraphics[width=80mm]{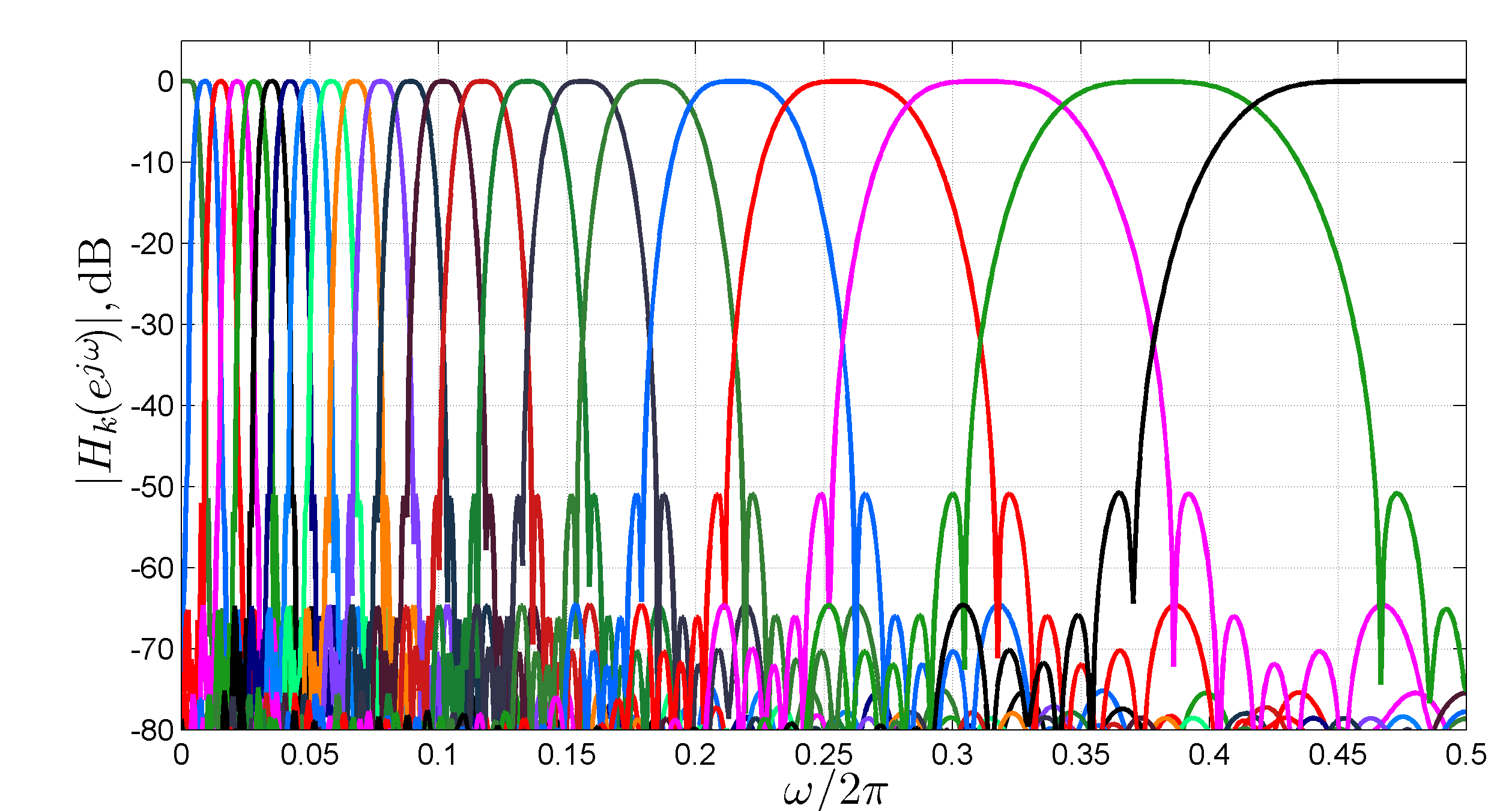}
\caption{Magnitude response of the warped 22-channel CMFB ($\alpha=0.5783$).}
\label{fig_5}
\end{figure}

The proposed design method was implemented using MATLAB on an Intel Celeron 2.8~GHz with 1 GB physical memory. It took nine outer loop iteration for the algorithm to converge (6 minutes). The frequency responses of initial and optimized filter prototypes are plotted in figure~\ref{fig_4}. It can be seen that proposed optimization procedure considerably minimized the stopband energy of filter prototype. Figure~\ref{fig_5} shows the resulting magnitude frequency response of resulting warped CMFB.

\begin{figure}[h]
\centering
\includegraphics[width=80mm]{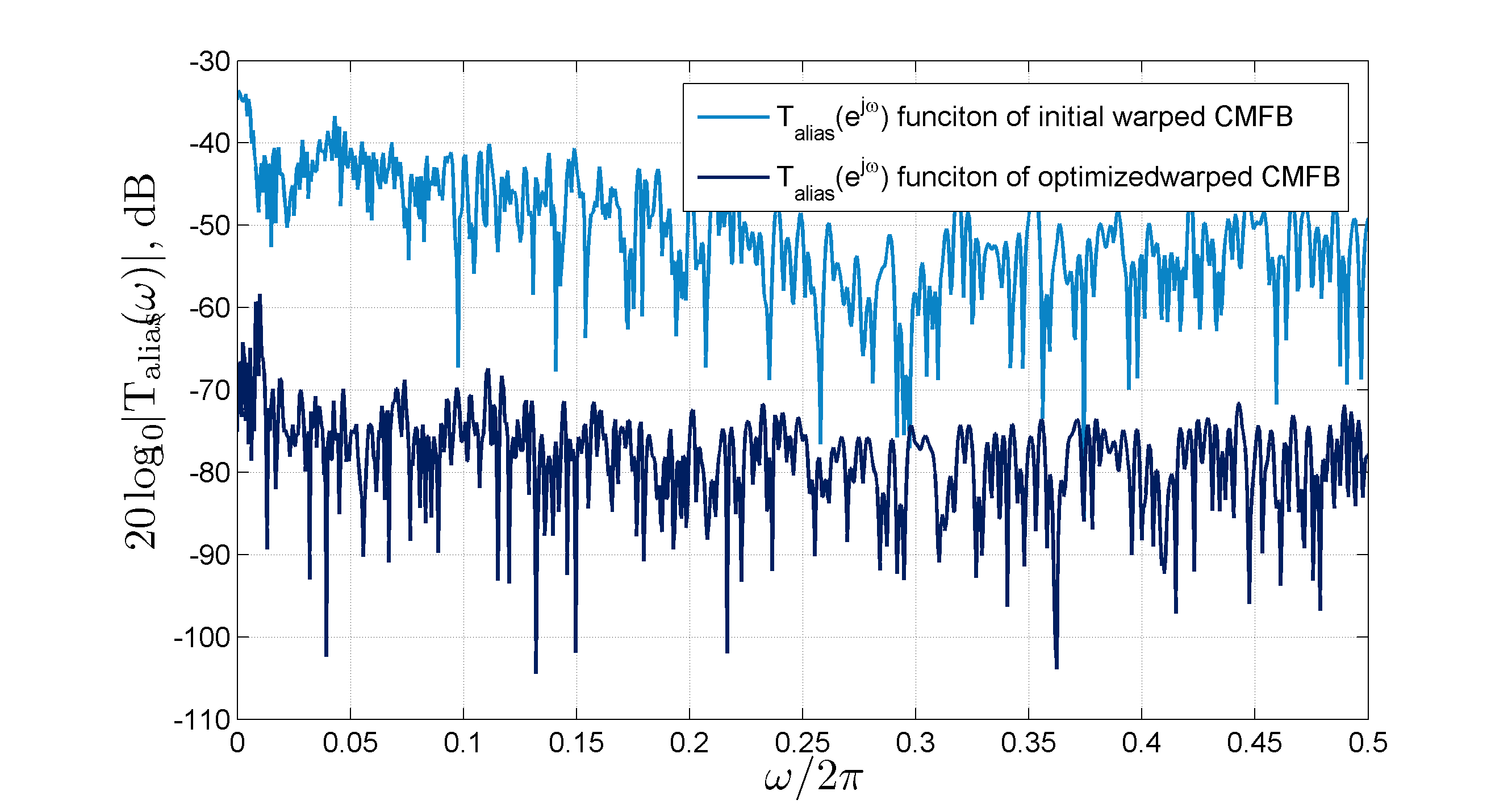}
\caption{Magnitude responses of aliasing transfer functions.}
\label{fig_6}
\end{figure}

For a chosen subsampling ratios $S_k$ the magnitude response of aliasing transfer functions for initial and optimized warped CMFB were calculated (figure~\ref{fig_6}). It can be seen that aliasing distortion has the same order of magnitude as the stopband attenuation of the prototype filters in figure~\ref{fig_4}.

\begin{figure}[htb]
\centering
\includegraphics[width=74mm]{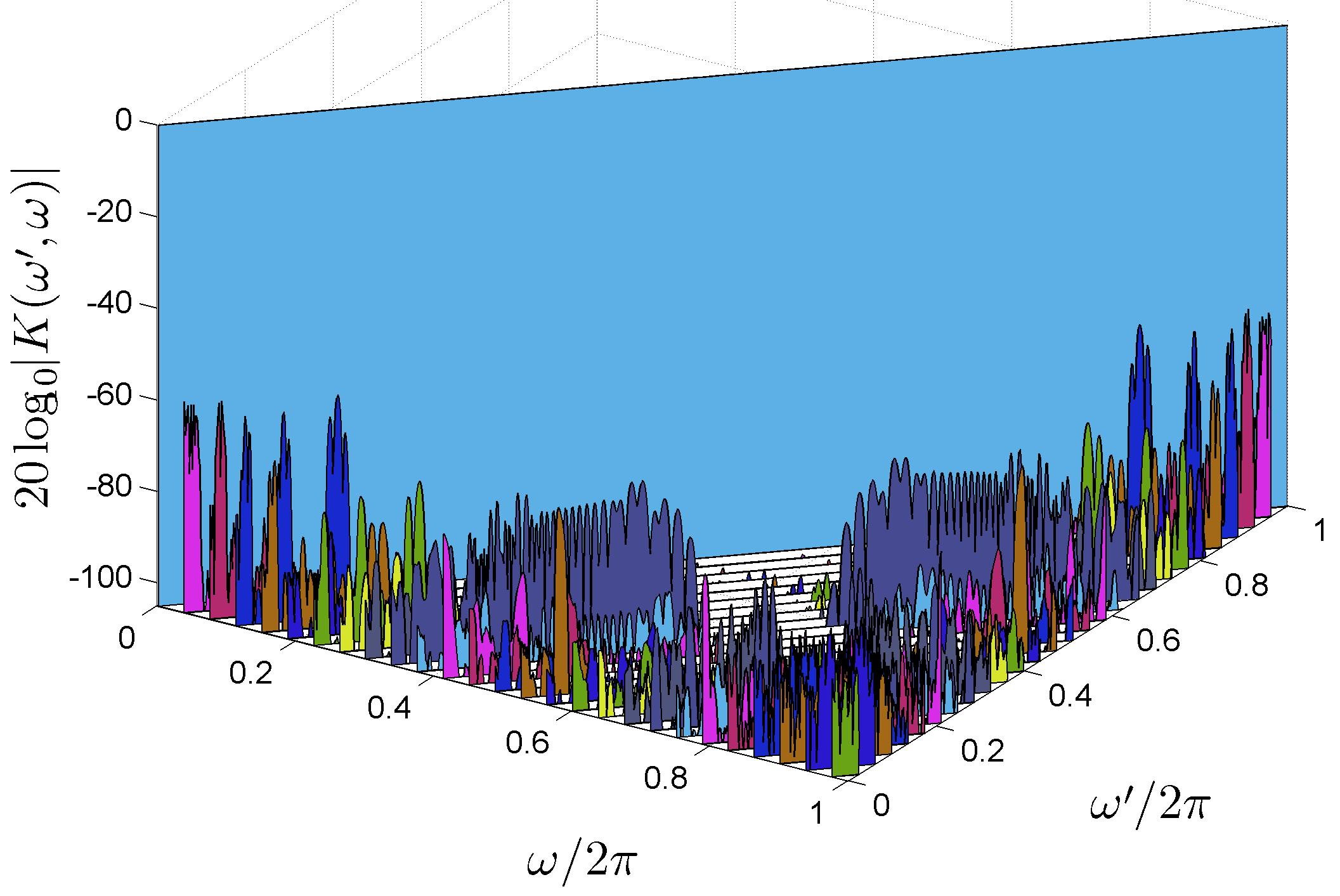}
\caption{Magnitude bifrequency response of initial warped CMFB.}
\label{fig_7}
\end{figure}

\begin{figure}[htb]
\centering
\includegraphics[width=70mm]{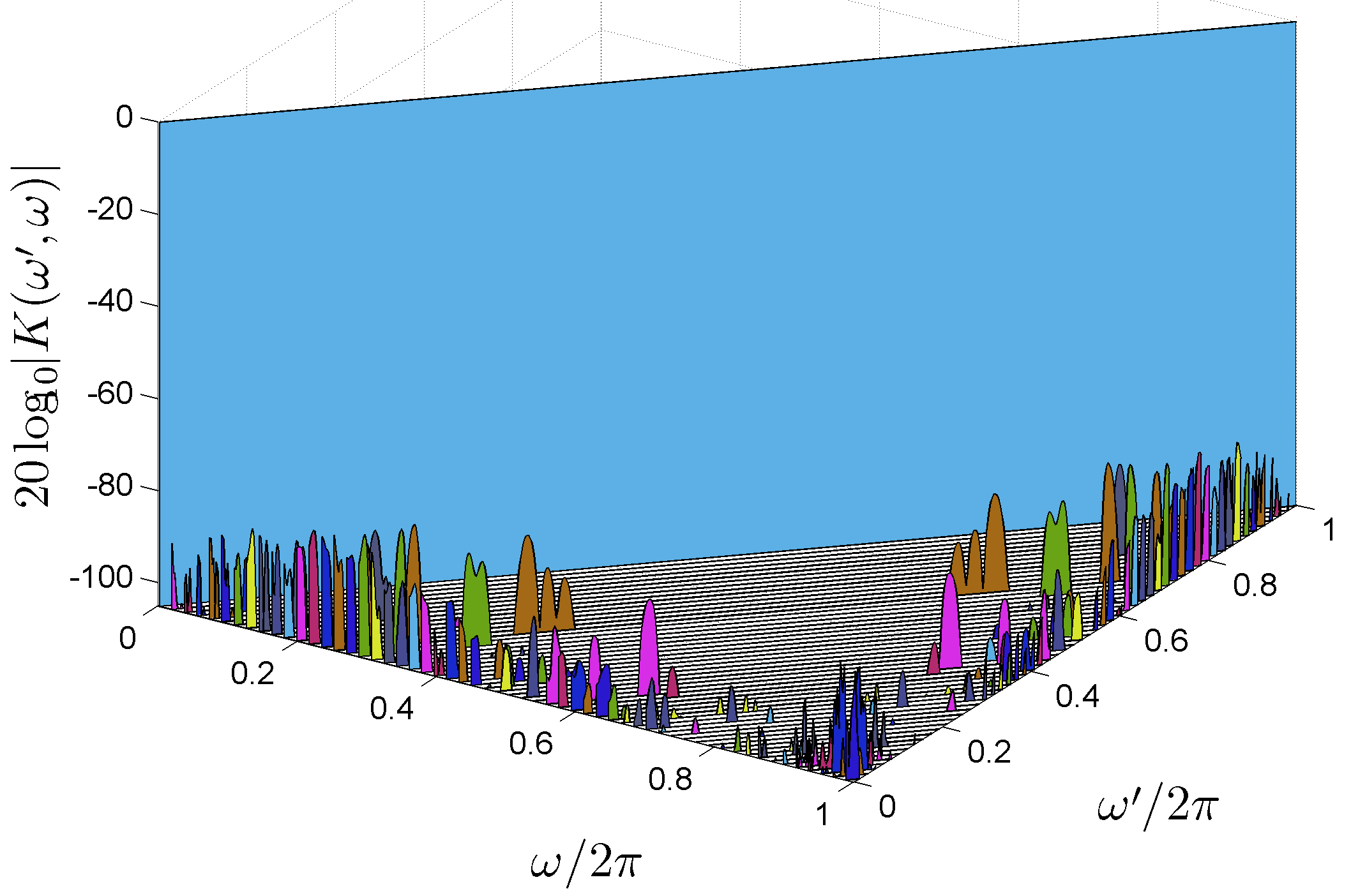}
\caption{Magnitude bifrequency response of optimized warped CMFB.}
\label{fig_8}
\end{figure}

\begin{figure}[htb]
\centering
\includegraphics[width=75mm]{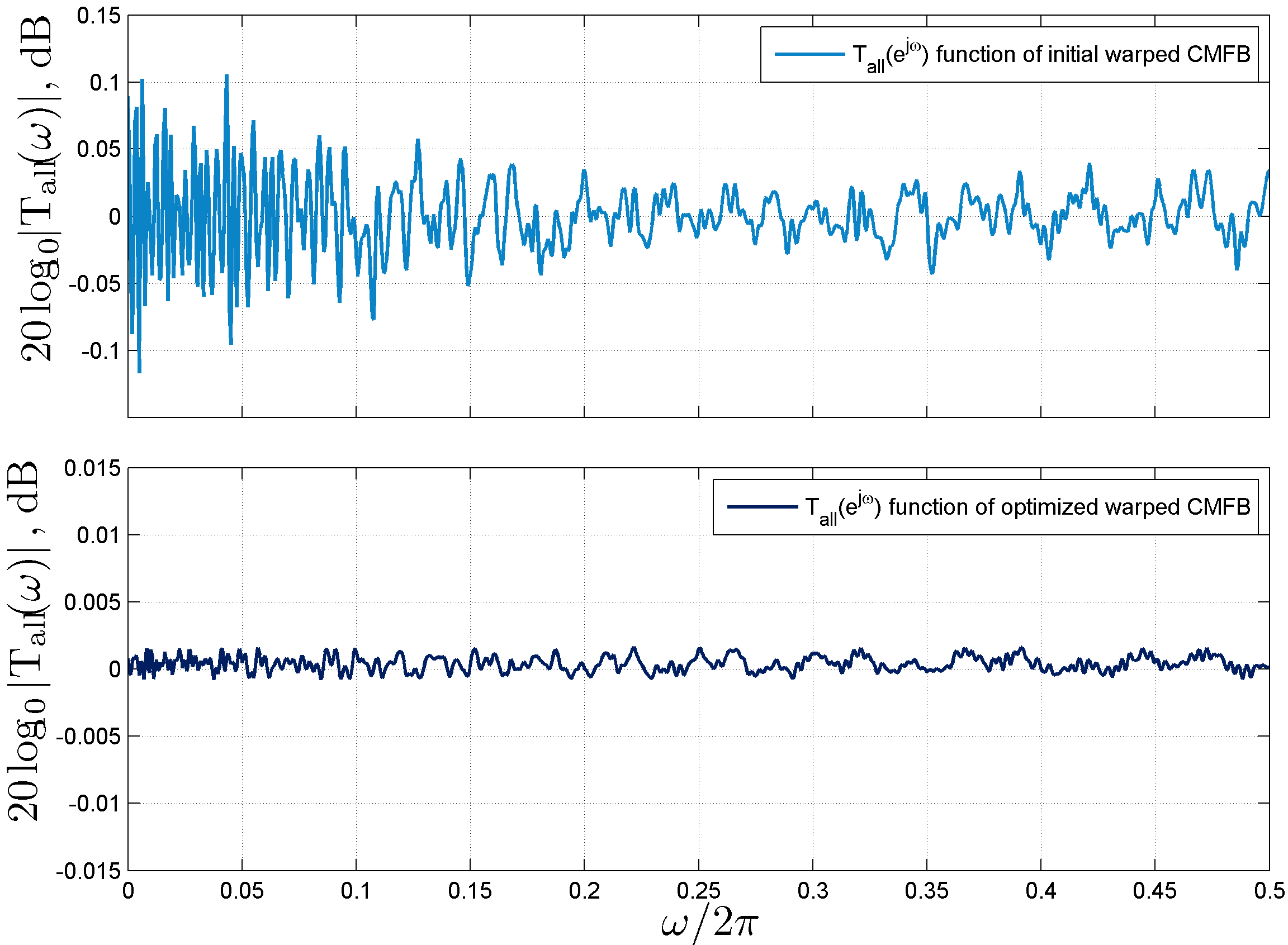}
\caption{Magnitude responses of $T_{all}(z)$.}
\label{fig_9}
\end{figure}

The level of aliasing component which appears due to decimation/interpolation of channel signals can be shown using bifrequency system function~\cite{01} (figure~\ref{fig_7}--\ref{fig_8}). Figure~\ref{fig_8} reveals that optimized warped CMFB attenuates aliasing component to the level of -80$\ldots$-90 dB.
Overall transfer functions of initial and optimized filter banks are given in figure~\ref{fig_9}. 
Overall transfer function of initial warped CMFB suffers from irregular distortion caused by aliasing. With optimized filter prototype the ripples of overall transfer are decreased significantly (from 0.15 dB to 0.004 dB). 
Thus the design example shows that proposed algorithm effectively minimizes the overall distortion introduced by warped CMFB.

The voice database and Matlab tools used for the voice analysis 

Matlab-implementation of proposed CMFB design method is available in public GitHub repository\footnote{https://github.com/Mak-Sim/Warped-filter-bank}.

\section*{\xiaosi\ \textbf{Conclusion}}
A practical method for the design of multichannel oversampled warped CMFB with low level of amplitude distortion has been proposed. Formulation of optimization problem and imposed constraints on overall filter bank transfer function are allowed to minimize amplitude distortion. Also the rule for selection of subsampling factor in warped CMFB is derived. Using the proposed design method it is possible to obtain high quality nonuniform filter bank with low distortion level.

\section*{\xiaosi\ \textbf{Acknowledgments}}

This work was supported by the Leading Academic Discipline Project of Shanghai Municipal Education Committee (J50104) and by the Belarusian Fundamental Research Fund (F11MS-037).

\bibliographystyle{IEEEbib}
\wuhao\

\wuhao\
\end{document}